\documentclass[aps,prb,twocolumn,superscriptaddress,showpacs,floatfix,longbibliography]{revtex4-2}
\usepackage{amsmath,amssymb,amsfonts,float,graphics,epsfig,epstopdf,color,verbatim,tabularx,bm,multirow,appendix,hyperref}
\usepackage[normalem]{ulem}
\usepackage{lmodern}
\usepackage{ytableau}
\usepackage{color}
\usepackage{bm}
\usepackage{wasysym}

\newcommand{\beq} {\begin{equation}}
	\newcommand{\eeq} {\end{equation}}
\newcommand{\bea} {\begin{eqnarray}}
	\newcommand{\eea} {\end{eqnarray}}
\newcommand{\be} {\begin{equation}}
	\newcommand{\ee} {\end{equation}}

\renewcommand{\)}{\right)}

\DeclareMathOperator{\Tr}{Tr}

\def\bq{{\mathbf{q}}}

\def\sign{{\text{sign}}}
\def\empty{{\text{empty}}}
\def\full{{\text{full}}}
\def\ii{{\text{i}}}

\begin{document}
	
\title {Fermion sign bounds theory in quantum Monte Carlo simulation}
\author{Xu Zhang}
\affiliation{Department of Physics and HKU-UCAS Joint Institute of Theoretical and Computational Physics, The University of Hong Kong, Pokfulam Road, Hong Kong SAR, China}
\author{Gaopei Pan}
\affiliation{Beijing National Laboratory for Condensed Matter Physics and Institute of Physics, Chinese Academy of Sciences, Beijing 100190, China}
\affiliation{School of Physical Sciences, University of Chinese Academy of Sciences, Beijing 100190, China}
\author{Xiao Yan Xu}
\email{xiaoyanxu@sjtu.edu.cn}
\affiliation{Key Laboratory of Artificial Structures and Quantum Control (Ministry of Education),
	School of Physics and Astronomy, Shanghai Jiao Tong University, Shanghai 200240, China}
\author{Zi Yang Meng}
\email{zymeng@hku.hk}
\affiliation{Department of Physics and HKU-UCAS Joint Institute of Theoretical and Computational Physics, The University of Hong Kong, Pokfulam Road, Hong Kong SAR, China}

\date{\today}
		
\begin{abstract}
Sign problem in fermion quantum Monte Carlo (QMC) simulation appears to be an extremely hard problem. Traditional lore passing around for years tells people that when there is a sign problem, the average sign in QMC simulation approaches zero exponentially fast with the space-time volume of the configurational space. We, however, analytically show this is not always the case and manage to find physical bounds for the average sign. Our understanding is based on a direct connection between the sign bounds and a well-defined partition function of reference system and could distinguish when the bounds have the usual exponential scaling, and when they are bestowed on an algebraic scaling at low temperature limit. We analytically explain such algebraic sign problems found in flat band moir\'e lattice models at low temperature limit. At finite temperature, a domain size argument based on sign bounds also explains the connection between sign behavior and finite temperature phase transition. Sign bounds, as a well-defined observable, may have ability to ease or even make use of the sign problem.
\end{abstract}

\maketitle

\section{Introduction}
Quantum Monte Carlo (QMC) is a standard and unbiased method for studying strongly correlated systems, widely used in condensed matter, high energy and quantum material research~\cite{Foulkes2001,Carlson2015,Assaad2008,Sandvik2010}. However, in reality, very often due to inadequate choices of basis, QMC simulations suffer from the so-called sign problem~\cite{Sugar1990exp,xu2022quantum,panSign2022}, in which configurational weights become negative or even complex and can no longer be interpreted as classical probabilities.
It has been proven that if the nondeterministic polynomial (NP) hard problem can be solved efficiently, then the obtained scheme can be used to solve the sign problem~\cite{NP-hard}, so that many interesting and fundamental questions in quantum many-body systems will be understood thereafter. Unfortunately, this has not happened yet.

Even though lacking a general guiding principle, many successful efforts have been done in finding systems without sign problem ~\cite{Lang1993,Koonin1997,Hands2000,2005CongjunWu,Chandrasekharan1999,Huffman2014,2015LeiWang,HongYao1,HongYao2,HongYao3,2016TaoXiang,Wei2017,Xu2019}, method to alleviate sign problem ~\cite{2015BasisChange,Rossi2017determinant,Rossi2017polynomial,DEmidio2020,2021BasisChange2,2020BasisChange3,2020BasisChange4,2020BasisChange5,2019BasisChange6,2020BasisChange7,Lef1,Lef2,Lef3,wan2020mitigating,2021Wynen,adiabatic2021}, origin of sign problem~\cite{Frustrated,XXZ,NP-hard,Hirsch1985,Topological_Origin,hastings2016quantum,ringel2017quantized,Intrinsic1,Intrinsic2} and connection of sign problem with phase transition~\cite{tarat2021,2021QPT,TCYi2021}. While there are some attempts for bosonic model in linking the average sign with the ratio of two partition functions\cite{Sugar1990exp,wesselThermodynamic2018,dEmidioReduction2020} at zero temperature limit, the requirement that two ground states are the same is too strict and sign problem is also either none or exponential. For fermionic system, it is even harder to find a system with the same decoupled space and ground states because of the involved determinant. Past experience and argument from those case by case efforts tells if there is a sign problem in QMC simulation, then the average sign will scale exponentially fast to zero with the space-time volume. But in our recent attempts of model design and QMC solution for flat band moir\'e lattice models at integer filling, both for real~\cite{ouyang2021projection} and momentum space~\cite{talk}, interesting outliers, where the average sign scales algebraically with system size and is independent with temperature at low temperature limit, were discovered. These results challenge  the conventional thoughts and urge a new understanding on universal properties of sign problem. 

To understand behavior of average sign, we suggest an observable called sign bounds which is basically the ratio of two related well-defined partition functions. With this tool, we may unveil bounds behavior of average sign analytically by studying ground state degeneracy (GSD) and ground state energy (GSE) and roughly estimate finite temperature phase transition point. 

We organize this article as below: First, we review determinant QMC and sign problem taking momentum space version as example. Then, we introduce sign bounds as an observable in sign bounds theory by suggesting two reference systems, and identify the situation where they can be used respectively by delivering two corollaries. We also give three cases related with flat band moir\'e materials~\cite{ouyang2021projection,XuZhang2021,JYLee2021,GaopeiPanValley2021,XuZhang2021SC} where average signs are all algebraic scaling with size at low temperature and the decay behavior of sign bounds is achieved exactly for any given size. Raising operator construction and tensor Young tableau method are used to derive the GSD analytically (see Appendix~\ref{sec:app2} and Appendix~\ref{sec:app3} for detail), which of course can be extended to count GSD for other SU(n) symmetry systems. And numerical QMC results for computing small size GSD as shown in Fig.~\ref{fig:fig3}(a) affirmed our derivation. Next, we point out sign bounds is also useful at finite temperature. Even though we can not derive this part analytically due to lacking detailed knowledge of partition function at finite temperature, an argument based on domain size can be used to distinguish finite temperature phase transition point roughly. Finally, we conclude our results and discuss the inspiration of sign bounds theory.

\section{Determinant QMC algorithm and sign problem}
First, we introduce the background that how determinant QMC works by following the momentum space version in Ref.~\cite{XuZhang2021}. The interaction Hamiltonian can be written as
\begin{eqnarray}
	\label{eq:eqmomentum1}
	H&=&\sum_{q\neq0} V(q)\rho_{q} \rho_{-q} \nonumber\\
	\rho_q&=&\sum_{k_1,k_2,m,n} \left( \lambda_{k_1,k_2,m,n}(q) c_{k_1,m}^\dagger c_{k_2,n} - \frac{1}{2} \mu_q\right)
\end{eqnarray}
Here, $V(q)$ is the Fourier transformation of any real space potential, $\lambda$ is the unitary transformation form factor of projecting interaction from plane wave basis to kinetic term diagonalized band basis. $\mu_q$ can be any constant number from projecting chemical potential terms. $k_1,k_2$ are used to label momentum and $m,n$ are used to label bands or any other freedom, e.g. spins or valleys. According to the discrete Hubbard-Stratonovich (HS) transformation, $e^{\alpha \hat{O}^{2}}=\frac{1}{4} \sum_{l=\pm 1,\pm 2} \gamma(l) e^{\sqrt{\alpha} \eta(l) \hat{o}}+O\left(\alpha^{4}\right)$, where $\alpha$ is a small constant number, $\hat{O}$ can be any operator, $\gamma(\pm 1)=1+\frac{\sqrt{6}}{3}$, $\gamma(\pm 2)=1-\frac{\sqrt{6}}{3}$, $\eta(\pm 1)=\pm \sqrt{2(3-\sqrt{6})}$ and $\eta(\pm 2)=\pm \sqrt{2(3+\sqrt{6})}$. We can rewrite the partition function as
\begin{widetext}
\begin{eqnarray}
	\label{eq:eqmomentum2}
	&Z&=\Tr\{\prod_{\tau}e^{-\Delta \tau H(\tau)}\} =\Tr\{\prod_{\tau} e^{-\Delta \tau \sum_{q\neq0} \frac{V(q)}{2} \left[\left(\rho_{-q}+\rho_{q}\right)^{2}-\left(\rho_{-q}-\rho_{q}\right)^{2}\right]  }\} \nonumber\\
	&\approx& \sum_{\{l_{|q|,\tau}\}} \prod_{\tau} [ \prod_{|q|\neq0}\frac{1}{16} \gamma\left(l_{|q|_1,\tau}\right) \gamma\left(l_{|q|_2,\tau}\right)]  \Tr\{\prod_{\tau}[\prod_{|q|\neq0}e^{i \eta\left(l_{|q|_1,\tau}\right) A_{q}\left(\rho_{-q}+\rho_{q}\right)} e^{\eta\left(l_{|q|_2,t}\right) A_{q}\left(\rho_{-q}-\rho_{q}\right)}]\}
\end{eqnarray}
\end{widetext}
Here $\tau$ is the imaginary time index with step $\Delta\tau$, $A_{q} =\sqrt{\frac{\Delta \tau V(q)}{2}}$ and $\{l_{|q|_1,\tau},l_{|q|_2,\tau}\}$ are the four-component auxiliary fields. By defining $P(\{l_{|q|,\tau}\})\equiv\prod_{\tau} [ \prod_{|q|\neq0}\frac{1}{16} \gamma\left(l_{|q|_1,\tau}\right) \gamma\left(l_{|q|_2,\tau}\right)]$ and ${D}(\{l_{|q|,\tau}\})\equiv\Tr\{\prod_{\tau}[\prod_{|q|\neq0}e^{i \eta\left(l_{|q|_1,\tau}\right) A_{q}\left(\rho_{-q}+\rho_{q}\right)} e^{\eta\left(l_{|q|_2,t}\right) A_{q}\left(\rho_{-q}-\rho_{q}\right)}]\}$, one can see partition function $Z$ as sample average for ${D}(\{l_{|q|,\tau}\})$ with configuration weight $P(\{l_{|q|,\tau}\})$.
\begin{equation}
	\label{eq:eqmomentum3}
	Z=\sum_{\{l_{|q|,\tau}\}} P(\{l_{|q|,\tau}\}) {D}(\{l_{|q|,\tau}\})
\end{equation}
Since $P(\{l_{|q|,\tau}\})$ here comes from continuous HS transformation coefficient (i.e., $\frac{1}{\sqrt{2\pi}} e^{-\frac{1}{2} x^2}$ in $e^{\alpha\hat{O}^2}=\frac{1}{\sqrt{2\pi}} \int e^{-\frac{1}{2} x^2}e^{\sqrt{2\alpha}x\hat{O}} dx$), it has already been normalized and only depends on configuration. Even though there is no sign problem for sampling this partition function as all configuration weights $P(\{l_{|q|,\tau}\})$ are non-negative, this sampling is extremely time consuming since one can see the weight here is just a Gaussian distribution and contains no information for the physical system. But it is still useful, e.g. determine small size GSD to benchmark the analytical method as shown in Fig.~\ref{fig:fig3}(a).

Now, we focus ourselves on how to compute an observable by QMC. Ensemble average of any observable $\hat{O}$ can be written as,
\begin{widetext}
\begin{equation}
	\langle \hat{O} \rangle = \frac{\Tr(\hat{O} e^{-\beta H})}{\Tr(e^{-\beta H})} = \sum_{\{l_{|q|,\tau}\}} \frac{P(\{l_{|q|,\tau}\}) \Tr[\prod_{\tau}\hat{B}_\tau(\{l_{|q|,\tau}\})] \frac{\Tr[\hat{O} \prod_{\tau}\hat{B}_\tau(\{l_{|q|,\tau}\})]}{\Tr[\prod_{\tau}\hat{B}_\tau(\{l_{|q|,\tau}\})]}}{\sum_{\{l_{|q|,\tau}\}} P(\{l_{|q|,\tau}\}) \Tr[\prod_{\tau}\hat{B}_\tau(\{l_{|q|,\tau}\})]} 
\end{equation}
Here $\hat{B}_\tau(\{l_{|q|,\tau}\})=\prod_{|q|\neq0}e^{i \eta\left(l_{|q|_1,\tau}\right) A_{q}\left(\rho_{-q}+\rho_{q}\right)} e^{\eta\left(l_{|q|_2,\tau}\right) A_{q}\left(\rho_{-q}-\rho_{q}\right)}$. Now, we see $W_l=P(\{l_{|q|,\tau}\}) \Tr[\prod_{\tau}\hat{B}_\tau(\{l_{|\bq|,\tau}\})]$ as possibility weight and $\langle \hat{O} \rangle_l=\frac{\Tr[\hat{O} \prod_{\tau}\hat{B}_\tau(\{l_{|q|,\tau}\})]}{\Tr[\prod_{\tau}\hat{B}_\tau(\{l_{|q|,\tau}\})]}$ as sample value for configuration $\{l_{|q|,\tau}\}$. Then Markov chain Mento Carlo can compute this $\langle \hat{O} \rangle$. That is how determinant QMC works in sign problem free case (i.e., $W_l$ is real and $W_l\geqslant0$ for all configurations). But this is not the case at most of time. With sign problem, one can still simulate according to QMC by reweighting~\cite{Sugar1990exp}
\end{widetext}

\begin{equation}
	\langle \hat{O} \rangle = \frac{\sum_{l} W_l\langle \hat{O} \rangle_l}{\sum_{l} W_l} = \frac{\frac{\sum_{l} |\Re(W_l)|\frac{W_l\langle \hat{O} \rangle_l}{|\Re(W_l)|}}{\sum_{l} |\Re(W_l)|}}{\frac{\sum_{l} W_l}{\sum_{l} |\Re(W_l)|}} \equiv \frac{\langle \hat{O} \rangle_{|\Re(W_l)|}}{\left\langle \sign \right\rangle }
\end{equation}

\begin{figure}
	\includegraphics[width=1.0\columnwidth]{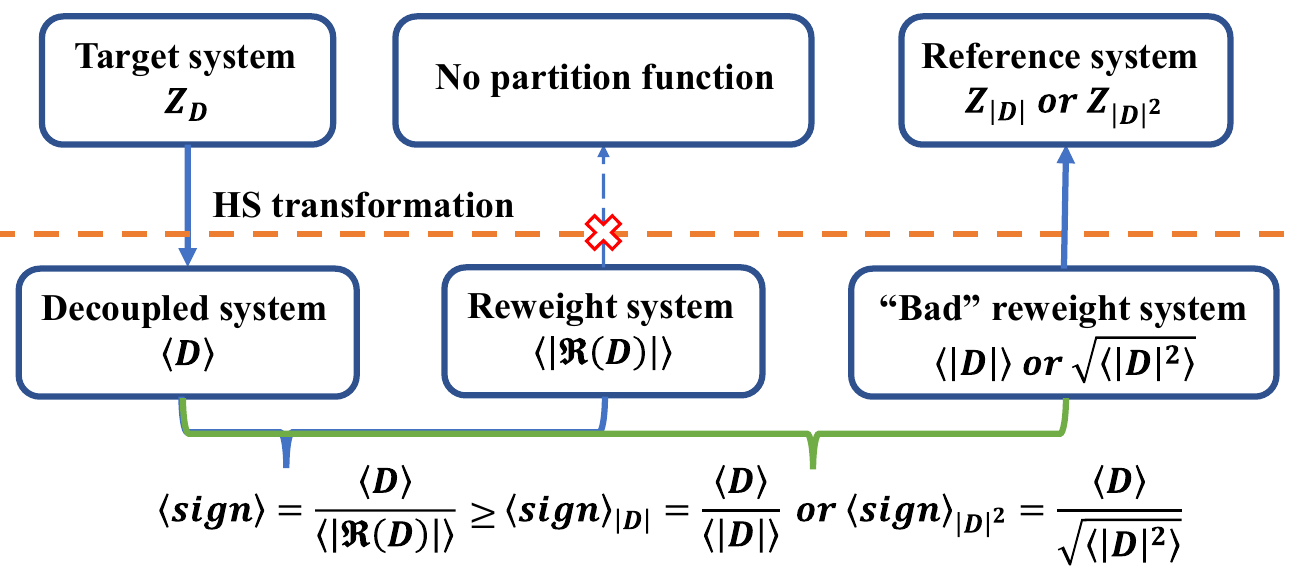}
	\caption{Schematical relationship between target system, reweight system and reference system. Average sign is ill-defined observable because reweight system can not recover to a physical system, while "bad" reweight system corresponds to a physical system so that sign bounds is a well-defined observable.}
	\label{fig:fig1}
\end{figure}

Here $\Re$ is taking real part operator. One can see $|\Re(W_l)|$ as a well-defined possibility weight and $\langle \hat{O} \rangle_{|\Re(W_l)|}$ is the measurement result according to this weight. The ratio $\frac{\sum_{l} W_l}{\sum_{l} |\Re(W_l)|}$ is the average sign $\left\langle \sign \right\rangle$ we keep talking about. One can see if $\left\langle \sign \right\rangle$ is exponentially small, fluctuation of computing $\langle \hat{O} \rangle$ from $\langle \hat{O} \rangle_{|\Re(W_l)|}$ will be exponentially large, which causes so called sign problem. To understand the sign behavior, one may interpret $\sum_{l} |\Re(W_l)|$ as a effective partition function~\cite{Sugar1990exp}. But most of time, this "partition function" is not well defined which means we can not write it back to a physical system before decomposition as shown in Fig.~\ref{fig:fig1}. This basically because we can hardly write the absolute value for real part of a determinant with another determinant. Then the idea comes out that if we can find a "bad" reweight system which can be written back to the partition function of a physical system (which we call a reference system) and there is a certain relationship between this well-defined partition function and the ill-defined "partition function" $\sum_{l} |\Re(W_l)|$, we find the bounds of $\left\langle \sign \right\rangle$ as a new physical observable which is defined as the ratio of two well-defined partition functions for related interactive systems. At low temperature, the exponential part with temperature can tell the GSE difference for those two systems, while the polynomial part with system size can tell GSD. Since we can measure both sign bounds and reference system, we can extract information for target system where there is a sign problem. At finite temperature if those two system have different phase transition behavior, e.g. 2D Ising and no finite temperature phase transition, this sign bounds can roughly tell the phase transition point according to a domain size argument. All of these will be discussed in detail below.

\begin{figure*}[t]
	\includegraphics[width=1.0\textwidth]{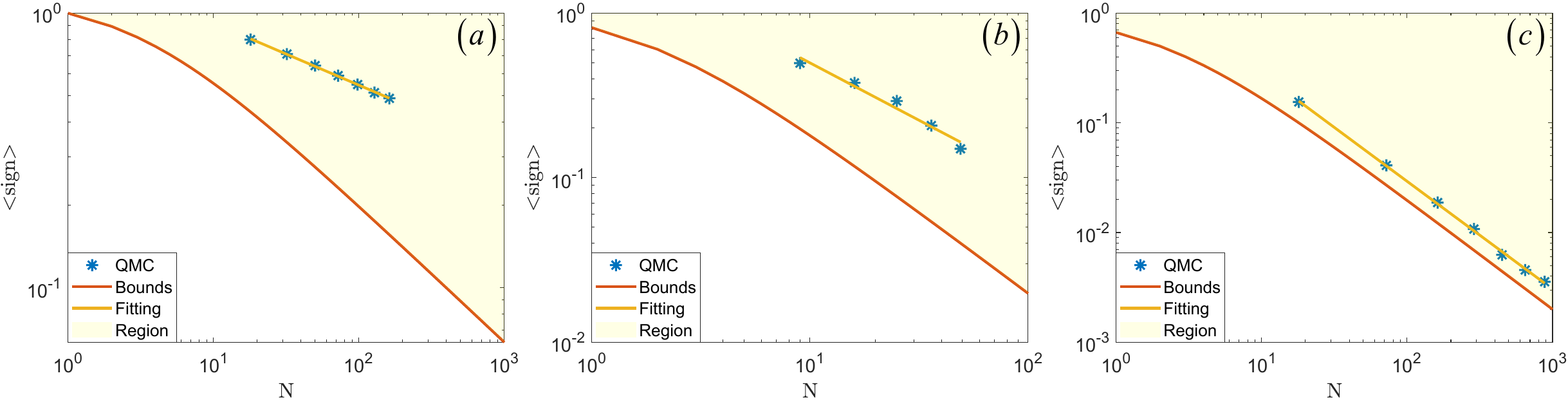}
	\caption{For three example cases, the average sign bounds, allowed region, QMC measurement $\left\langle \sign \right\rangle$ and its polynomial fitting with fitting lines are shown. The errorbars in QMC data are smaller than the symbol size. (a) For momentum space case 1 model, bounds is determined by $\left\langle \sign\right\rangle \geqslant {2}/{\sqrt{N+3}}$. QMC measurements for $N=18,32,50,72,98,128,162$ are carried out at a $\left\langle \sign \right\rangle$ converged temperature. Fitting line $\propto$ $N^{-0.23}$. (b) For momentum space case 2 model, bounds is determined by $\left\langle \sign\right\rangle \geqslant 2 / {\sqrt{(N+1)^2+2}}$. QMC measurements for $N=9,16,25,36,49$ are carried out at $T=0.91$ meV. Fitting line $\propto N^{-0.70}$. (c) For real space model, bounds is determined by $\left\langle \sign\right\rangle \geqslant 2/(N+2)$ and $N=18,72,162,288,450,648,882$ are measured from QMC~\cite{ouyang2021projection}. Fitting line $\propto N^{-0.98}$.}
	\label{fig:fig2}
\end{figure*}

\section{Sign bounds theory}
For a generic quantum many-body system with Hamiltonian $H$, we have partition function $Z = \Tr\(e^{-\beta H}\)$. Then as shown in \eqref{eq:eqmomentum3}, the partition function can be written as the weighted average of fermion determinants after the Trotter decomposition and HS transformation, $Z_{D} = \sum_{\{l\}} P(\{l\}) D(\{l\}) \equiv \left\langle D\right\rangle $, where $l$ represents different configurations, $P(\{l\})$ is the configurational weight for determinant $D(\{l\})$. Since we only consider Hermitian system by default, $Z_{D}=\Re(Z_{D})= \left\langle \Re(D)\right\rangle$. Now, imagine we can find a "bad" reweight system with the same configuration space. As a reweight system, it requires non-negative determinant $V(\{l\}) \geqslant 0$ for any configuration $l$. Besides, we also require $\left\langle V\right\rangle  \equiv\sum_{\{l\}} P(\{l\}) V(\{l\})$ corresponds to a partition function $Z_V$ of a physical Hamiltonian before HS decomposition. We call this physical Hamiltonian the reference system $V$ and define the average sign bounds based on $V$ as $\left\langle \sign \right\rangle_V$. This sign bounds can give bounds behavior of $\left\langle \sign \right\rangle$ based on the inequality between $\left\langle |\Re(D)|\right\rangle$ and $f(\left\langle V\right\rangle)$. Here $f$ can be any function of $\left\langle V\right\rangle$. For a concrete understanding, we give two reference systems $V=|D|^2$, $V=|D|$ which satisfies $\left\langle D\right\rangle=\left\langle \Re(D)\right\rangle \leqslant \left\langle |\Re(D)|\right\rangle \leqslant\sqrt{ \left\langle |D|^2\right\rangle }$ and $\left\langle D\right\rangle=\left\langle \Re(D)\right\rangle \leqslant \left\langle |\Re(D)|\right\rangle \leqslant \left\langle |D|\right\rangle$. One can see those reference systems do give bounds as
\begin{eqnarray}
\label{eq:eqsignbounds1}
\frac{Z_D}{\sqrt{Z_{|D|^{2}}}} \equiv \frac{\left\langle D\right\rangle}{\sqrt{ \left\langle |D|^2\right\rangle }} \leqslant \left\langle \sign\right\rangle \equiv \frac{\left\langle D\right\rangle}{\left\langle |\Re(D)|\right\rangle}  \leqslant 1 \nonumber\\
\frac{Z_{D}}{Z_{|D|}} \equiv \frac{\left\langle D\right\rangle}{\left\langle |D|\right\rangle} \leqslant \left\langle \sign\right\rangle \equiv \frac{\left\langle D\right\rangle}{\left\langle |\Re(D)|\right\rangle}  \leqslant 1
\end{eqnarray}
The logic is drawn schematically in Fig.\ref{fig:fig1}. For now, the non-negative requirement and inequality between $\left\langle |\Re(D)|\right\rangle$ and $f(\left\langle V\right\rangle)$ are both trivial which means it is true for any system. Then the only non-trivial requirement for a reference system is it must have a physical partition function so that we can say sign bounds is a well-defined observable. Below we give two corollaries where two possible reference systems $Z_{|D|^2}$, $Z_{|D|}$ can apply separately on a large group of systems.

{\it{Corollary I.-}}
For a fermion Hamiltonian whose sign of QMC for every configuration is real, we can introduce another U(2) freedom $s\in\{+,-\}$ such as spin or valley, to prepare a reference system.

This is what we usually do to avoid sign problem when designing a sign-problem-free model to study general effects of certain interaction. But here we do not simulate this sign-problem-free system as our target system. We extract physical information of our target system with sign problem from sign bounds behavior $\frac{Z_D}{\sqrt{Z_{|D|^{2}}}}$. For example, if this ratio does not change with temperature when temperature goes to zero, the GSE $E$ between reference system and target system must have the certain relation $E_D=E_{|D|^{2}}/2$. And the GSD $g$ can be derived from $\frac{Z_D}{\sqrt{Z_{|D|^{2}}}}=\frac{g_D}{\sqrt{g_{|D|^{2}}}}$. Here $g_{|D|^{2}}$ is the GSD with this U(2) freedom. For a global discrete symmetry of the Hamiltonian, the ground states belong to finite-dimension irreducible representation which contributes degeneracy independently with system size. For global continuous symmetry such as $SU(n)$, degeneracy is determined by normal Young diagram ground states belong to, which is polynomial with system size. Then polynomial average sign behavior with system size will be seen. We will discuss this zero GSE and global continuous symmetry case below.

{\it{Examples.-}}
We introduce two cases here, which are set in momentum space, with a generic PSD Hamiltonian (e.g., describing the long-rangle Coulomb interaction in flat band system)
\begin{equation}
	H=\sum_{q\neq0} V(q) \rho_{-q} \rho_q = \sum_{q\neq0} V(q) \rho_q^\dagger \rho_q
	\label{eq:eq1}
\end{equation}
where $\rho_q=\sum_{i,j} \left( \lambda_{i,j}(q) c_i^\dagger c_j - \frac{1}{2} \mu_q\right) $, $i,j$ are matrix indexes with dimension $N$ (e.g., the momentum grid in moir\'e Brillouin zone, $N=6\times6$, $9\times9$, $\cdots$), and the averagely half-filled physical system requires $V(q)=V(-q)>0$, $\Tr(\lambda_{i,j}(q))=\mu_q$. In our previous momentum space QMC work~\cite{XuZhang2021}, we proved that $\Tr(\lambda_{i,j}(q))=\mu_q$ guarantees the sign for every configuration is real. Then one can require zero-energy ground state for observing $\left\langle \sign \right\rangle \geqslant g_{D}/\sqrt{g_{|D|^{2}}}$ behavior.

\begin{figure*}[t]
	\includegraphics[width=1.0\textwidth]{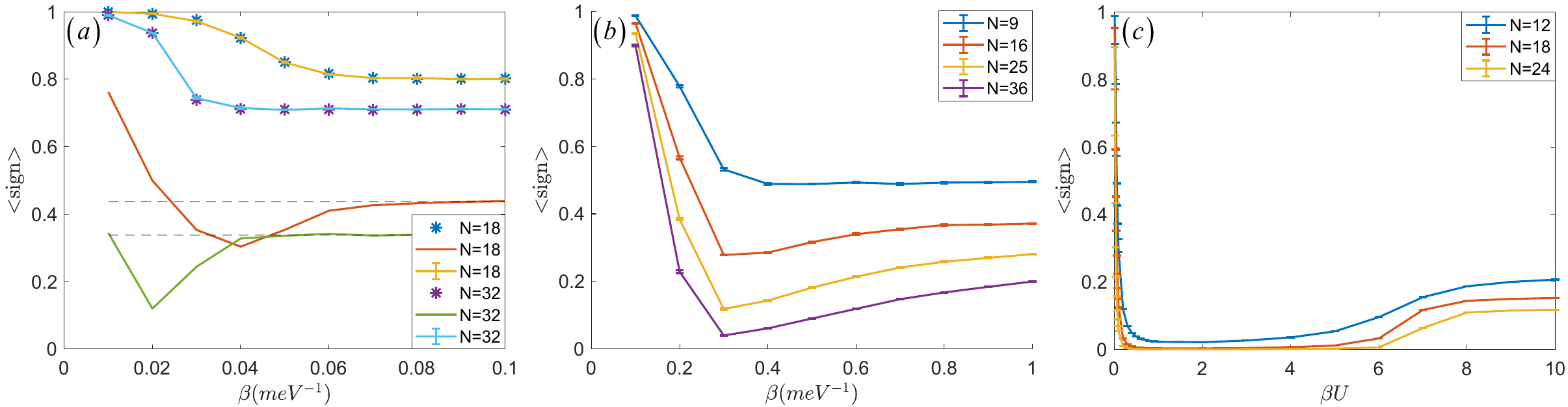}
	\caption{Finite temperature results for three example cases. (a) For momentum space case 1 model, size $N=18,32$ are plotted. We measure $\left\langle \Re(D)\right\rangle$, $\left\langle |\Re(D)|\right\rangle$ and $\sqrt{ \left\langle |D|^2\right\rangle }$ according to $Z=\sum_{\{l\}} P(\{l\}) {D}(\{l\})$. Stars are $\left\langle \sign\right\rangle$ determined by $\frac{\left\langle \Re(D)\right\rangle}{\left\langle |\Re(D)|\right\rangle}$ and lines without error bar are sign bounds determined by $\frac{\left\langle \Re(D)\right\rangle}{\sqrt{ \left\langle |D|^2\right\rangle }}$, dashed lines are low temperature limit bounds $\frac{2}{\sqrt{N+3}}$. Error bar lines are $\left\langle \sign\right\rangle$ computed independently with importance sampling QMC. (b) For momentum space case 2 model, we plot $\left\langle \sign\right\rangle$ computed with importance sampling QMC for size $N=9,16,25,36$. (c) For real space case 3 model, we plot $\left\langle \sign\right\rangle$ computed with importance sampling QMC for size $N=12,18,24$.}
	\label{fig:fig3}
\end{figure*}

In our case 1, $\mu_q=0$ and $\lambda_{i,j}(q)$ is set randomly. For $\mu_q=0$, one can easily check empty and full states are two ground states with zero energy. Random $\lambda_{i,j}(q)$ means there should be no other symmetry so that no other degeneracy. Then we have $g_D=2$. After we introduce another freedom $s\in\{+,-\}$, which means now $\rho_{s,q}=\sum_{i,j} \left( \lambda_{i,j}(q) \cdot \left( c_{i,+}^\dagger c_{j,+} + c_{i,-}^\dagger c_{j,-}\right)  -  \mu_q\right) $. For computing degeneracy $g_{|D|^2}$, one can define a raising operator
\begin{equation}
	\Delta^\dagger=\sum_{i'}c_{i',+}^\dagger c_{i',-}.
\end{equation}
By noticing $\left[ \Delta^\dagger, \rho_{s,q}\right]=0$ and $\left| \psi_{+,\empty}\right\rangle \otimes \left| \psi_{-,\full}\right\rangle$ is a ground state. We can apply $\Delta^\dagger$ at most $N$ times on this state, and this gives us $N+1$ ground states with $N$ particles. Besides, we also have ground states with $0$ and $2N$ particles (i.e. $\left| \psi_{+,\empty}\right\rangle \otimes \left| \psi_{-,\empty}\right\rangle$ and $\left| \psi_{+,\full}\right\rangle \otimes \left| \psi_{-,\full}\right\rangle$). Together, $g_{|D|^2}=N+3$ for random $\lambda_{i,j}(q)$ so that $\left\langle \sign \right\rangle \geqslant {g_D}/{\sqrt{g_{|D|^2}}} = {2}/{\sqrt{N+3}}$ as shown in Fig.~\ref{fig:fig2}(a). This means at low temperature and large size limit, $\left\langle \sign \right\rangle$ will decay no faster than $N^{-\frac{1}{2}}$. It is worth to mention this derivation and the tensor Young tableau method below should be {\it exact} and can be checked for small size numerically as shown in Fig.~\ref{fig:fig3}(a) unless there is accidental degeneracy.

In our case 2, $\mu_q\neq0$ and there are two band labels $m,n\in\{1,-1\}$ for $\lambda_{i,j,m,n}(q)=m\cdot n\cdot\lambda_{i,j,m,n}(q)$ (e.g., chiral flat band limit for spin-polarized and valley-polarized TBG at half filling, following gauge chosen in Ref.~\cite{bernevig2020tbg3,lian2020tbg4}). Now $\rho_q=\sum_{i,j,m,n} \left( \lambda_{i,j,m,n}(q) c_{i,m}^\dagger c_{j,n} - \frac{1}{2} \mu_q\right)$ and there are two degenerate ground states with Chern number $\pm 1$, i.e. $g_D=2$. Again, after introducing freedom $s\in\{+,-\}$, $\rho_{s,q}=\sum_{i,j,m,n} \left( \lambda_{i,j,m,n}(q)\cdot \left( c_{i,m,+}^\dagger c_{j,n,+} + c_{i,m,-}^\dagger c_{j,n,-}\right)  - \mu_q\right)$. One can define two raising operators (see Appendix~\ref{sec:app2})
\begin{eqnarray}
	\Delta_1^\dagger&=&\sum_{j'} (c_{j',1,+}^\dagger +\ii c_{j',-1,+}^\dagger) (c_{j',1,-} -\ii c_{j',-1,-}) \nonumber\\
	\Delta_2^\dagger&=&\sum_{j'} (c_{j',1,+}^\dagger -\ii c_{j',-1,+}^\dagger) (c_{j',1,-} +\ii c_{j',-1,-})
\end{eqnarray}
Here $\ii$ is the imaginary unit. It is straightforward to verify $[\Delta_1^\dagger, \rho_{s,q}]=[\Delta_2^\dagger, \rho_{s,q}]=0$ and $[\Delta_1^\dagger, \Delta_2^\dagger]=[\Delta_1^\dagger, \Delta_2]=0$. This means $\Delta_1^\dagger$ and $\Delta_2^\dagger$ generate two groups of orthogonal eigenstates with the same energy. By noticing $\left| \psi_{+,\empty}\right\rangle \otimes \left| \psi_{-,\full}\right\rangle$ is a ground state, apply $\Delta_1^\dagger$ or $\Delta_2^\dagger$ independently will give $(N+1)^2$ orthogonal zero-energy states with zero Chern number. Here $N$ is dimension of label $i,j$ in $\lambda_{i,j,m,n}(q)$. Besides, there are also two states with non-zero Chern number, on which raising operators apply are equal to 0 so that no other states will be given, $g_{|D|^2}=(N+1)^2+2$. This means for this model, $\langle \sign \rangle \geqslant {2}/{\sqrt{(N+1)^2+2}}$ as shown in Fig.~\ref{fig:fig2}(b).

{\it{Corollary II.-}}
For a Hamiltonian with a PSD interaction part $\sum_{A} (A-\mu_A)^2$ and a kinetic part $K$, where $A$ and $K$ are the fermion bilinears and $\mu_A$ is real constant number. If for a certain group of $\mu_A$, there is no sign problem, then $Z_{|D|}$ can be used as partition function of reference system (This can be seen by noticing $\mu_A$ only contributes a phase in $D(\{l\})$, so that the reference system now is just the Hamiltonian with the sign-problem-free $\mu_A$). At low temperature, $\left\langle \sign \right\rangle \geqslant g_D e^{-\beta (E_D - E_{|D|})}/g_{|D|} $. If $E_D = E_{|D|}$, $\left\langle \sign \right\rangle \geqslant g_D/g_{|D|}$.

{\it{Example.-}}
It is easy to notice this corollary can give a sign bounds behavior for all kinds of repulsive Hubbard model (by seeing $\mu_A$ as chemical potential or other on-site ionic potential), where the usual exponential decay of average sign is commonly seen. Though QMC may say little about $E_D$ and $g_D$, one could expect with the improvement of constrained path QMC, density matrix renormalization group and tensor-network type methods, GSE and GSD for finite size doped Hubbard system can be obtained 
~\cite{White_1993,DMRG_RMP,RMP_mps,MPQin2020,BBC2021}, so that theoretical sign bounds behavior $g_D e^{-\beta (E_D - E_{|D|})}/g_{|D|}$ can be achieved and check with the numerical one. Since there have been many discussions on the sign problem therein~\cite{Rubtsov2005,assaad1990reinvestigation,Sugar1990exp,Furukawa1991,Scalettar2015,Kim2020,tarat2021}, here we still focus the application of this reference system on a PSD Hamiltonian without kinetic part $K$. We study the Kang-Vafek's real space model~\cite{Kang2019} for twisted bilayer graphene at flat band limit with $1/4$ or $3/4$ filling ($\nu=\pm2$), whose GSE $E=0$ is equal to that at half filling ($\nu=0$). In this example, we give an explanation for algebraic average sign observed in Ref.~\cite{ouyang2021projection} and determine the sign bounds analytically. The model is written below
\begin{equation}
	H=U\sum_{\varhexagon} \left( Q_{\varhexagon}+\alpha T_{\varhexagon}-\nu\right) ^2
\end{equation}
where $Q_{\varhexagon}=\frac{1}{3}\sum_{\sigma,\tau}\sum_{l=1}^{6} c^\dagger_{R+\delta_l,\sigma,\tau} c_{R+\delta_l,\sigma,\tau} - 4$,  $T_{\varhexagon}=\sum_{\sigma,\tau}\sum_{l=1}^{6} \left[ (-1)^l c^\dagger_{R+\delta_{l+1},\sigma,\tau} c_{R+\delta_l,\sigma,\tau} + h.c.\right] $, $\nu$ is used to control filling, $\sigma,\tau$ are spin and valley indexes, $R+\delta_{l}$ represents site $l$ in a single $R$ hexagon and $U, \alpha$ are real constants. Attention to the subtraction of a constant $4$ in the definition of operator $Q_{\varhexagon}$ which moves the charge neutrality point to $\nu=0$. We consider the infinite-$U$ case as studied in Ref.~\cite{ouyang2021projection}. Fixing $\alpha$, then the only parameter is $\beta U$, and any finite $\beta$ in the infinite-$U$ limit corresponds to low temperature limit. We consider the system with linear system size $L$ up to $L=21$ and total number of sites is $N=2L^2$. This model is sign problem free at charge neutrality~\cite{PhysRevX.11.011014,Liao_2021}, with partition function identical to $Z_{|D|}$, as when $\nu$ is away from charge neutrality, it only introduce a phase to the weight~\cite{ouyang2021projection}. As the real space model is PSD, and one can construct zero energy ground state both at charge neturality and at $\nu = \pm 2$ with GSD $g_D=(N+3)(N+2)(N+1)/6$ for $\nu=\pm 2$~\cite{Kang2019,lian2020tbg4} and GSD $g_{|D|}=(N+3)(N+2)^2(N+1)/12$ for $\nu=0$~\cite{lian2020tbg4} (see Appendix~\ref{sec:app3} for the Tensor Young tableau method in which the GSD are obtained), therefore $\left\langle  \sign \right\rangle  \geqslant g_D/g_{|D|} = 2/(N+2)$. Fig.~\ref{fig:fig2}(c) shows the average sign presented in Ref.~\cite{ouyang2021projection}, comparing with the sign bounds we proposed.

\section{Sign bounds at finite temperature}
We will show how sign bounds can roughly imply finite temperature phase transition point with a domain size argument in this section.

It is worthwhile to notice when we are talking about low temperature limit at finite temperature and finite size system simulation, we are actually considering the simulated system size is covered by a single domain within which the partition function can be well described by GSE and GSD. That is the low temperature limit we discussed above. For finite temperature case, we generally can not achieve partition function explicitly. But we expect partition functions for target system and reference system should have different behaviors with temperature according to different symmetry or irreducible presentation their ground states belong to.

Take our first two cases as example, one can see the target system has $Z_2$ symmetry for double degenerated ground states while reference system has continuous $SU(2)$ symmetry with continuous excitations. For 2D system, we know a $Z_2$ symmetry ground state has a finite temperature phase transition, while there is no finite temperature phase transition for $SU(2)$ symmetry. This difference causes a result that, when increasing temperature from zero, partition function for target system $Z_D$ is almost unchanged because the system is still in a single ground state domain as long as below phase transition temperature, while continuous excitations from $SU(2)$ symmetry will contribute to $Z_V$ so long as the temperature fluctuation suppress this domain size below the system size. This is the first step shown in Fig.~\ref{fig:fig3}(b), more states contribute to $Z_V$ will make $\left\langle \sign\right\rangle$ decrease. Things change when energy scale of temperature touches excitation breaking $Z_2$ symmetry. The correlation length of ground state begins decrease linearly with inverse of temperature $\tau^{-1}$ which is faster than $SU(2)$ symmetry. It means domain size decreases faster for target system. At this time, excitations contribute more to $Z_D$ than $Z_V$. This is the second step of in Fig.~\ref{fig:fig3}(b), $\left\langle \sign\right\rangle$ begins increase after phase transition of target system. Finally, fluctuation from temperature almost makes all freedom of Hamiltonian contribute equally to partition function. That is high temperature limit where $\left\langle \sign\right\rangle\approx1$ is well understood.

According to this argument, one can forecast the excitation gap should enlarge with size for our case 1, since the lowest point of sign bounds shift to high temperature with increasing size as shown in Fig.~\ref{fig:fig3}(a). This behavior is unphysical just because we use random matrix to construct our Hamiltonian. In our physical case 2, different size dip at similar temperature which indicates the phase transition temperature roughly. Besides, one can also forecast the domain size of target system decrease slower than reference system at low temperature in our case 3 so that there is a reduction of $\left\langle \sign\right\rangle$ when increasing temperature as as shown in Fig.~\ref{fig:fig3}(c), even though target system and reference system here almost have the same symmetry. Stars and error bar line match well in Fig.~\ref{fig:fig3}(a) also indicates the way we compute $Z=\sum_{\{l\}} P(\{l\}) {D}(\{l\})$ is valid for finite size at finite temperature. And the converging to dash line at low temperature limit shows our GSD analysis is correct.

\section{Conclusion and discussion}
We bridge a connection between the sign problem and well-defined partition function properties of the target system and reference system in interactive quantum many-body problem. At low temperature, the sign bounds depends on GSE and GSD of the target system and the reference system. At finite temperature, a domain size argument based on symmetry is delivered. Our two corollaries give two groups of systems where reference systems ${|D|^2}$, ${|D|}$ can apply separately so that sign bounds theory can work. As applications of our two corollaries at low temperature limit, we demonstrate the algebraic scaling of average sign in a class of PSD models which have physical background with flat band quantum mori\'e materials. The sign bounds theory we proposed is a generic criteria that could help with finding new quantum many-body models that acquire the algebraic sign or mild exponential sign.

Inspiration idea can also be extended to projection version of QMC~\cite{Ceperley1980,Sugiyama1986,Sorella1989a,Sorella1988numerical,Zhang1995}. Suppose we have a trial wavefunction with $|\Psi_T\rangle = \sum_n c_n |\Psi_n\rangle$ where $|\Psi_n\rangle$ is the eigenstate of $H$ with eigenvalue $E_n$. In the projection version of QMC, the true wavefunction is achieved by projection, i.e., $|\Psi_0\rangle \sim \lim_{\Theta \rightarrow \infty } e^{-\Theta H}|\Psi_T\rangle$. To keep similar notation with finite temperature case, we denote $Z_D = \langle \Psi_T | e^{-2\Theta H}|\Psi_T\rangle = c_D e^{-2\Theta E_D}$, where $c_D=\sum_{n\in D} |c_n|^2$ is the ground state component in $|\Psi_T\rangle$ and $E_D$ is the GSE. For reference system, we have $Z_V=c_V e^{-2\Theta E_V}$, where $c_V$ is the reference system ground state component in $|\Psi_T\rangle$ and $E_V$ is the reference system GSE. Then the sign bounds is
\begin{equation}
\langle \sign \rangle_V = \frac{c_D e^{-2\Theta E_D}}{c_V e^{-2\Theta E_V}}.
\end{equation}
This implies to alleviate the sign problem, it is important to make the trial wavefunction have larger components of the ground state and make the reference system have closer GSE with origin system, though this generally still can not remove exponential part from energy difference. This may explain the recent proposed adiabatic method, where the trial wavefunction is adiabatically improved and the closeness of reference GSE is well expected~\cite{adiabatic2021}.

In a similar spirit, with the further improvement of DMRG and tensor-network type of approaches, where the low temperature, GSE and GSD on finite size systems are expected to be obtained with better accuracy, it will be natural to make use of these information and perform the analysis with our theory to provide (hopefully improved) bounds of QMC on the systems with conventional exponential sign problem, such as doped Hubbard model or quantum spin systems with frustration. Moreover, as our theorem directly relates the sign problem to the physical properties of the system, it implies there may be important information hidden inside configurations even when the simulation has sign problem. Therefore, machine learning and related data mining of Monte Carlo data with sign problem are encouraging~\cite{broecker2017machine}.

\begin{acknowledgments}
	{\it Acknowledgments}\,---\, X.Y.X. thanks stimulating discussions with T. Grover on related topics. We thank Zhiwen Zhang for useful discussions on the mathematical perspectives of our theory. X.Y.X. is sponsored by the National Key R\&D Program of China (Grant No. 2021YFA1401400), Shanghai Pujiang Program under Grant No. 21PJ1407200, Yangyang Development Fund, and startup funds from SJTU.
	X.Z., G.P.P. and Z.Y.M. acknowledge support from the RGC of Hong
	Kong SAR of China (Grant Nos. 17303019, 17301420, 17301721
	and AoE/P-701/20) and the Strategic Priority Research
	Program of the Chinese Academy of Sciences (Grant
	No. XDB33000000), the K. C. Wong Education Foundation (Grant No. GJTD-2020-01) and the Seed Funding "Quantum-Inspired
	explainable-AI" at the HKU-TCL Joint Research Centre for Artificial Intelligence. We thank the Computational Initiative at the Faculty of Science and the Information Technology Services at the University of Hong Kong and the Tianhe platforms at the
	National Supercomputer Center in Guangzhou for their
	technical support and generous allocation of CPU time.
\end{acknowledgments}

\newpage
\begin{widetext}

\appendix
\section{Parameters for three cases}
\label{sec:app1}
For the first two cases, we take $V(q)$ in the equation of main text as single gate screened Coloumb potential used in Ref.~\cite{XuZhang2021}
\begin{equation}
	V(\bar{q}) \approx 7.925 \frac{1}{\sqrt{N_{k}}\bar{q}}\left(1-e^{-22.36 \frac{1}{\sqrt{N_{k}}} \bar{q}}\right) meV
\end{equation}
Here, $\bar{q}$ is the distance between momenta in moir\'e Brillouin zone (mBZ) by setting two nearest $k$ points with unit length and $N_{k}$ is the number of momentum points in mBZ. We take a cut-off for $\bar{q}$ that the longest $\bar{q}$ is the length of reciprocal lattice vectors of the mBZ. $m,n$ in $\lambda_{i,j,m,n}$ are two flat band indexes for TBG. Form factor $\lambda$ is the overlap between two flat band wave functions at momentum $k_1,k_2$ (i.e., $\lambda_{k_1,k_2,m,n}=\sum_{G',X}u^{*}_{k_1,m;G',X}u_{k_2,n;G',X}$). For case 1, we just take this $2N_{k}\times2N_{k}$ matrix $\lambda_{k_1,k_2,m,n}$ randomly with requiring $\Tr(\lambda)=0$ and maximum mode 1 for every complex element. For our case 2, we compute flat band wave functions at chiral limit with BM continuous model ~\cite{Trambly2010,Trambly2012,Bistritzer_TBG,ROZHKOV20161,Santos2007,Santos2012} parameters $(\theta,\hbar v_F/a_0,u_0,u_1)=(1.08^{\circ},2.37745eV,0eV,0.11eV)$. For convenience, we put BM continuous model Hamiltonian below
\begin{equation}
	H^{\tau}_{k,k^\prime}=\left(\begin{array}{cc} 
		-\hbar v_F (k-K_1) \cdot \sigma\delta_{k,k^\prime}  &  U_0\delta_{k,k^\prime}+U_1 \delta_{k,k^\prime+G_1}+U_2 \delta_{k,k^\prime+G_1+G_2}  \\
		U_0^\dagger\delta_{k,k^\prime}+U_1^{\dagger} \delta_{k,k^\prime-G_1}+U_2^{\dagger} \delta_{k,k^\prime-G_1-G_2}  & -\hbar v_F (k-K_2) \cdot \sigma\delta_{k,k^\prime}
	\end{array}\right)
\end{equation}
Here, the explicit $2\times2$ matrix is labeled by layer index, and in each layer there is also a implied $2\times2$ matrix labeled by sublattice index. $\sigma$ is just Pauli matrix for describing Dirac cone of monolayer graphene. Twist angle $\theta$ is set as magic angle where two flat bands emerge. The $\theta$ will be included in Hamiltonian by noticing moir\'e lattice constant is determined by $a_M=a_0/\theta$ when $\theta$ is small. Interlayer tunneling between two Dirac cones located at $K_1$ and $K_2$ in the same valley $\tau$ is described by matrixes $U_0=
\left(\begin{array}{cc} 
	u_0  & u_1 \\ 
	u_1 & u_0
\end{array}\right)$, $U_1=\left(\begin{array}{cc} 
	u_0  & u_1 e^{-i\frac{2\pi}{3}} \\ 
	u_1 e^{i\frac{2\pi}{3}} & u_0
\end{array}\right)$ and $U_2=\left(\begin{array}{cc} 
	u_0  & u_1 e^{i\frac{2\pi}{3}} \\ 
	u_1 e^{-i\frac{2\pi}{3}} & u_0
\end{array}\right)$
where $u_0$ and $u_1$ are the intra-sublattice and inter-sublattice interlayer tunneling amplitudes. $a_0$ is the lattice constant of monolayer graphene, $G_1$ and $G_2$ are reciprocal vectors of mBZ and $\tau$ in $H^{\tau}_{k,k^\prime}$ means one only considers one valley $\tau$ Hamiltonian and ignores inter-valley tunneling.

For our case 3, the model parameters used in Fig.2 are the same to the parameters used in Ref.~\cite{ouyang2021projection}. We set $\alpha=0.5$, and only have $\beta U$ as an independent parameter. We divide $\beta U$ into $L_\tau$ slices $\beta U = L_\tau \Delta_\tau U$, where $L_\tau$ scales with $L$, $L_\tau=10L$. We then use the infinite-$U$ projection scheme, which correponds to zero temperature limit. Therefore, we have perfect scaling of average sign in Fig.~\ref{fig:fig2}(c). In Fig.~\ref{fig:fig3}, we calculate system size $3\times 2$, $3\times 3$ and $3\times 4$, as each unit cell has two sites, we have totoal number of sites $N=12$, 18 and 24. We do not allow $L_\tau$ to scale with $L$ in this case, but determine it through $L_\tau = \beta U/\Delta_\tau U$, where $\beta U$ ranges from 0.01 to 10 (for $\beta U$ from 0.01 to 0.09, $\Delta_\tau U=0.001$ is used, for $\beta U$ from 0.1 to 1, $\Delta_\tau U=0.01$ is used, for $\beta U$ from 2 to 10, $\Delta_\tau U=0.1$ is used), and the infinite-$U$ projection scheme is not used, therefore it correponds to a finite temperature study.

\section{Raising operator construction}
\label{sec:app2}
Here we introduce the way we use to look for raising operators. First, for random $\lambda_{i,j}$, $\rho_q=\sum_{i,j} \left( \lambda_{i,j}(q) c_i^\dagger c_j - \frac{1}{2} \mu_q\right) $ and $\rho_{s,q}=\sum_{i,j} \left( \lambda_{i,j}(q) \cdot \left( c_{i,+}^\dagger c_{j,+} + c_{i,-}^\dagger c_{j,-}\right)  -  \mu_q\right) $. If we want to find a two-fermion operator $\Delta^\dagger$ commuting with $\rho_{s,q}$, we only need to find a matrix commuting with $\rho_{s,q}$ in single-particle basis. The total dimension of $\rho_{s,q}$ can be written into $i,j$ space direct products $s\in\{+,-\}$ space noted by $K\otimes S$. Since generally $\lambda_{i,j}$ in $K$ space does not have any symmetry, one need a unit operator in this space to commute with $\rho_{s,q}$. While in $S$ space, one can see $\rho_{s,q,i,j}$ looks like
\begin{equation}
	\rho_{s,q,i,j}=
	\left(\begin{array}{cc} 
		\lambda_{i,j}(q)  & 0 \\ 
		0 & \lambda_{i,j}(q)
	\end{array}\right)
\end{equation}
which means actually any $2\times2$ matrix will commute with $\rho_{s,q,i,j}$. We choose a raising operator which makes charge flip from $-$ to $+$ space.
\begin{equation}
	\Delta^\dagger_{i,j}=
	\left(\begin{array}{cc} 
		0  & 1 \\ 
		0 & 0
	\end{array}\right)
\end{equation}
Direct product those two parts of $\Delta^\dagger$, we achieve the raising operator we want 
\begin{equation}
	\Delta^\dagger=\sum_{i'}c_{i',+}^\dagger c_{i',-}
\end{equation}
Then with similar spirit, we can derive $\Delta^\dagger_1$, $\Delta^\dagger_2$ in case 2. The only difference is now we have another band space ${1,-1}$. At chiral limit, $\rho_{s,q,i,j,+,-}$ in band space looks like
\begin{equation}
	\rho_{s,q,i,j,+,-}=
	\left(\begin{array}{cc} 
		\lambda_{i,j,1,1}(q)  & \lambda_{i,j,1,-1}(q) \\ 
		-\lambda_{i,j,1,-1}(q) & \lambda_{i,j,1,1}(q)
	\end{array}\right)=\lambda_{i,j,1,1}(q)\sigma_0 + \ii \lambda_{i,j,1,-1}(q) \sigma_y
\end{equation}
Here $\sigma_0$ is a unit matrix and $\sigma_y$ is the Pauli matrix. For commuting with this matrix, the $2\times2$ matrix can be only in the form of $\alpha \sigma_0 + \beta \sigma_y$. For now, any choosing of $\alpha,\beta$ can generate one reasonable raising operator. If one also would like two different raising operators with different choosing of $\alpha,\beta$ satisfy $[\Delta^\dagger_1, \Delta_2]=0$, which means they are two groups of independent operators commuting with each other, $(\alpha_1 \sigma_0 + \beta_1 \sigma_y)(\alpha_2 \sigma_0 + \beta_2 \sigma_y)=0$ must be satisfied. One simple choice is $(\sigma_0 + \sigma_y)(\sigma_0 - \sigma_y)=0$. Then we achieve $\Delta^\dagger_1, \Delta^\dagger_2$ we used in the case 2 in the main text,
 \begin{eqnarray}
	\Delta_1^\dagger&=&\sum_{j'} (c_{j',1,+}^\dagger +\ii c_{j',-1,+}^\dagger) (c_{j',1,-} -\ii c_{j',-1,-}), \nonumber\\
	\Delta_2^\dagger&=&\sum_{j'} (c_{j',1,+}^\dagger -\ii c_{j',-1,+}^\dagger) (c_{j',1,-} +\ii c_{j',-1,-}).
\end{eqnarray}

\section{Tensor Young tableau method}
\label{sec:app3}
For self consistency, we introduce another useful way, which is also used in Ref.~\cite{lian2020tbg4,PhysRevB.75.184441,JYLee2021} , to compute the ground state degeneracy besides raising operator construction. This method works for models whose degeneracy are introduced by SU(n) symmetry. $g_D=(N+3)(N+2)(N+1)/6$ for $\nu=\pm 2$ and GSD $g_{|D|}=(N+3)(N+2)^2(N+1)/12$ for $\nu=0$ in Kang-Vafek's real space model are calculated in this way.

In Kang-Vafek's real space model case, degeneracy is introduced by SU(4) symmetry, where spin and valley contribute two SU(2). One can easily see any m-particle wave function can always be described by a 4-dimension rank-m tensor like $T_{a_1,a_2,...,a_m}$, where different $a_i\in\{1,2,3,4\}$ with 4 spin-valley dimensions label particles. SU(4) transformation for spin-valley space and $S_m$ permutation for particle index are independent and will only change the tensor $T$ to another tensor $T'$ in the tensor space, which means $T$ is well-defined for decomposition according to irreducible representation. By noticing the number of the same $a_i$ must be less than lattice size $N$, the Young diagram must have rows less than 4 and columns less than $N$. The Young diagram to which full-filling one spin-valley flavor  ground state for $\nu=\pm 2$ belongs is
\begin{center}
	$\underbrace{\ydiagram{4}\cdots\cdots\ydiagram{4}}$ \\
	N
\end{center}
Assume only this irreducible representation contributes to ground states. Then counting the degeneracy is equal to count normal Young tableau of this Young diagram, which can be calculated by hook's rule

\begin{equation}
	d_{\left[ N\right] }(SU(4)) = \prod_{j}\frac{4+j-1}{j} = \frac{\frac{(N+3)!}{3!}}{N!} = \frac{(N+3)(N+2)(N+1)}{6}
\end{equation}

While the Young diagram to which full-filling two spin-valley flavors  ground state for $\nu=0$ belongs is just
\begin{center}
	$\underbrace{\ydiagram{4,4}\dots\dots\ydiagram{4,4}}$ \\
	N
\end{center}

The number of normal Young tableau calculated by hook's rule is

\begin{equation}
	d_{\left[ N,N\right] }(SU(4)) = \prod_{i,j}\frac{4+j-i}{h_{i,j}} = \frac{\frac{(N+3)!(N+2)!}{3!2!}}{(N+1)!N!} = \frac{(N+3)(N+2)^2(N+1)}{12}
\end{equation}

The numerical way to compute the GSD by QMC according to $Z=\sum_{\{l\}} P(\{l\}) {D}(\{l\})$ also confirms the assumption that only one irreducible representation contributes to ground states in this case.

\end{widetext}

\bibliographystyle{apsrev4-2}
\bibliography{SignProblem.bib}

\end{document}